# Scheduling Algorithms for Asymmetric Multi-core Processors

Alan David

*Abstract*— Growing power dissipation due to high performance requirement of processor suggests multicore processor technology, which has become the technology for present and next decade. Research advocates asymmetric multi-core processor system for better utilization of chip real state. However, asymmetric multi core architecture poses a new challenge to operating system scheduler, which traditionally assumes homogeneous hardware. So, scheduling threads to core has become a major issue to operating system kernel. In this paper, proposed scheduling algorithms for asymmetric multicore processors have been categorized. This paper explores some representative algorithms of these classes to get an overview of scheduling algorithms for asymmetric multicore system.

I. INTRODUCTION

To meet the high performance requirement of processor, chip manufacturers has been historically relaying on decreased chip size and increased transistor number they contain. The technological advancement of complementary metal–oxide–semiconductor (COMS) has enabled manufacturers to create densely integrated chip operating at higher frequency, thus meet the goal of ongoing increased high performance challenge. According to Moore's law, transistors, that can be placed inexpensively on integrated circuit doubles every two year. But ironically, the transistor technology has also posed power barrier this days so that, transistors can not be shrunk continuously and make a more dense integrated circuit due to the fact that power consumption and power dissipation has also increased. For example, Pentium-4 class processors consume 50W and processors of 2015 are expected to consume around 300W [1]. Thus, processor performance increment has begun slowing. According to Linley Group president Linley Gwennap, in 1990s chip performance increased 60 percent per year but it slowed down to 40 percent per year from 2000-2004 [2].

Growing concern for increased power dissipation suggests alternate processor technology and multi-core processor is a new step forward, which has become the technology for current and next decade [3, 4]. Multi-core processors combine two or more independent cores in single die. Thus, a Dual core processor means two cores in a die; a Quad core means four cores in a die and so on. Multi-core processors take advantage from the fundamental relationship between frequency and power. Each core in Multi-core processors runs in a lower frequency compare to single processor. Power is also divided among cores of Multi-core processors, giving high performance gain with decreased average power dissipation in a core. Another difference between single processors and Multi-core processors is that single processors have unique L-1 and L-2 cache where in Multi-core processors each core has independent unique L-1 cache but cores share the L-2 Cache. So, without no doubt, manufacturers have turned on for Multi-core processors. IBM first introduced Multi-core processor chip, power 4 in 2001 [5]. Using this, designers were able to achieve higher performance and greater communication bandwidth. Intel also gained energy-efficient performance in mid -2006 with their Intel CoreTM2 Duo processors [6]. Now questions arise, should all cores in Multi-core processors necessarily be identical/ homogeneous/ symmetric or different/ heterogeneous/ asymmetric? Should the instruction set also be different among cores?

Recent research [7, 8, 9, 10, 11, 12] advocates asymmetric multi-core processor for better utilization of chip real estate where, a processor can have multiple cores with different performance characteristics. An asymmetric multi-core processor would contain cores that have the same instruction-set architecture, but they would differ in features, size, performance, power consumption etc. Core diversity offers higher value than uniformity for many applications, offering the greater opportunity to adapt with the requirement of the application [10]. System software such as a scheduler can pick a core according to requirement of the application while runtime and thus, can lead energy-efficient computing. Homogeneous multi-core processors cannot take the opportunity of resource requirement of applications. For example, applications that has large amount of ILP (instruction-level parallelism) can be executed in a core that has the ability to exploit ILP (a core that issue multiple instructions per second, a wide-issue super scalar CPU). This core will optimize power consumption of the processor. But, this core certainly will not be a good choice for the applications which have little amount of ILP, resulting wastage of power compare to the other simpler core that has the low power consumption characteristics due the ability to handle applications with less ILP. If all cores are similar (just simply copy of a core) then there is no way to exploit resource constraint feature of application. To argue for the asymmetric Multi-core processors, Kumar [10] et al shows that 39% average energy reduction possible only sacrificing 3% of performance if the objective function is used that optimized energy efficiency with a tight performance threshold and in case of an objective function that optimizes for energy delay with looser performance bound, the result is nearly three times improvement on energy-delay product while sacrificing only 22% performance.

Operating system scheduler traditionally assumes homogeneous hardware. Thus, asymmetric multi-core processors creates an unique challenge to kernel, as scheduling threads in a "good" asymmetric aware manner although gives the benefit of heterogeneity, the "bad" way may loose all benefits not only for heterogeneity but also the benefits of using multiple cores. Balakrishnan et al. [13] shows that an asymmetry unaware scheduler would not only result in bad application performance, but can also cause instability of application.

## II. RELATED WORK

Most works on scheduling for asymmetric Multi-core processors have done in operating system level. The asymmetry-aware algorithms [12, 15, 16, 17] schedule threads on cores through continuous monitoring of performance and analysis. Another algorithm [18] schedules thread relying on information about application. Fair scheduling was a characteristic to optimize in traditional system. Balakrishnan et al. [8] describes the fair scheduling algorithms ensuring fair sharing of the fast cores. Restricted form of asymmetry, where cores run at different frequencies, has been discussed by a number of research papers [19, 20].

On the other hand, Winter et al. [37] conclude that prior global power management algorithms based on linear programming requires high computational requirement and they are not feasible for many cores. They proposed an algorithm called Steepest Drop achieves orders of magnitude lower execution time without sacrificing power-performance efficiency.

Vahid et al. implement changes that are required to make a hypervisor scheduler to asymmetry-aware scheduler at [22]. In [23], impact of asymmetry in distributed memory system has been discussed. The authors assumed static assignment policies to assign thread to processors and further assumed many nodes with smaller caches for support multi thread parallelism, few nodes with large caches to support single thread parallelism. Kumer et al. presents resource sharing between adjacent cores as a means for saving area of the die and the overall improvement of system performance [24]. In [25], a method of synthesizing custom architecture is proposed, where asymmetry is ensured by augmenting the instruction of symmetric processor through custom instructions.

Another type of core specialization was proposed by Kumar et el, where a slow core runs the controlling domain of a virtual machine monitor Xen [26]. In similar vein, Mogul et al. proposed an algorithm where slow cores were reserved for executing system calls [27].

Thread level parallelism (TLP) specialization algorithms consider sequential and parallel application type while assigning jobs to asymmetric processors. To optimize energy, fast cores can be used to execute sequential code where slow cores may be employed to execute parallel code. Algorithm in [28] address TLP while thread scheduling. To combine the both types of specialization, CAMP [16], a scheduling algorithm has been proposed by Juan et al., which delivers both kind of specialization (efficiency and TLP). They introduced a metric Utility Factor (UF), which gives a single value depending upon the efficiency and TLP of the application. This value implies how much application will improved its performance if all of its threads are allowed to occupy the fast cores of the system.

Nagesh et al [30] proposed a new policy named as aged scheduling policy where the algorithm predicts the remaining execution time of threads based on their age and assigns a thread to fast core, which has larger remaining execution time.

## III. PAPER FOCUS

Thread scheduling is one of the fundamental services offered by the operating system kernel. Some of the characteristics that scheduler wants to optimize are fairness, response time, turnaround time, throughput and efficiency [14]. For a symmetric Multi-core processors, all cores are identical, the kernel scheduler still can optimize feature described earlier keeping the load of each core in consideration. But, for a scheduler working on an asymmetric Multi-core processors will have to not only consider the workload of the cores, but also it has to consider the relative performance of computation among the cores. For example, scheduler in Multi-core processors architecture should assign threads demanding high CPU (computationally complex) to the cores that are "fast" [12]. Fast core may have high clock frequency, complex super-scalar out-of-order pipelines, improved branch-prediction and pre-fetching hardware. On the other hand, threads demanding low CPU but high memory intensiveness may be scheduled in "slow" cores. Slow cores are characterized by the opposite of fast cores, say, lower operating frequency, less complex hardware and simple in-order pipeline. Slow cores will occupy less area than fast cores and surely consume less power. In systems containing large number of slow cores and few number of fast cores, by scheduling memory intensive or high stall prone threads on slow processors will result less power consumption of energy without any significant performance loss relative to fast core. So, it is clear that, traditional scheduling policies of single processor or homogeneous Multi-core architecture will not allow to take optimization of using asymmetric Multi-core processor. Here comes the scope for new scheduling algorithms to take a part in asymmetric Multi-core processor architecture.

As discussed earlier section, scheduling algorithms can be categorized into three categories- efficiency specialization algorithms, thread level parallelism specialization algorithms and the algorithms that exploit both. Efficiency specialization algorithms try to achieve improved efficiency by assigning the most CPU intensive threads to fast cores. The algorithms discussed in [12, 15, 29] are some representative algorithms of this class. Thread level parallelism specialization algorithm try to get improved performance by assigning sequential applications and sequential phases of parallel application to fast cores. Algorithm discussed in [28] is a representative algorithm of this class. Algorithm in [16] delivers both type of specialization. In this paper we will discuss these algorithms so that we get a comprehensive overview of all the types of algorithm.

## IV. EFFICIENCY SPECIALIZATION-HASS SCHEDULER

In [29], Daniel et al explain an algorithm that schedules threads based on some signatures of threads collected offline. The scheduler is called HASS (Heterogeneity aware signature supported) scheduler. This scheduler is based on architectural signatures of threads, which can be defined as the architectural properties of threads. An architectural signature of an application may consists of available instruction level parallelism of the application, memory-boundedness, sensitivity of various clock speed and other parameters. In short it is a compact summary of architectural properties of the application. The properties of signatures give the scheduler an idea or way of finding proper matching for an application to a core. HASS scheduler maps application to cores based on the properties of application signature. This signature is populated in offline fashion and served to scheduler as a single unit with the application binary.

### A. What is Signature?

The summary of architectural characteristics of an application is called signature. If the load and characteristics of a core is provided then HASS scheduler depends on the ability to predict potential performance of a thread on a core. So, the signature characteristics should be complete enough to give the scheduler the opportunity to predict threads' relative performance on different cores. The paper focuses on systems where cores differ in clock frequency and cache size.

Memory-boundedness of an application should be also taken care of while considering performance variation due to clock frequency. An application with high level of frequent memory access will stall the core, resulting negative consequence on performance. So, the authors consider estimation of cache miss rates is the contents of the signature. Reuse- distance profile which is the distribution of the number of intervening memory access between consecutive accesses to the same memory location, is an indication of memory-boundedness of an application. Last level cache miss rate can be estimated using reuse-distance profile.

### B. How Signatures will be constructed?

As the scheduler at scheduling time should need the signature, so application binary can be used to hold the signature. To construct the signature first reuse-distance profile is needed. Offline profiling can be used to collect reuse-distance profile. Feedback-directed optimization phase of the application development with no or little involvement from the developer may be an example of offline profiling. After collecting offline-profiling cache misses of limited set of realistic cache configuration is measured which comprise the architectural signature.

### C. How Signatures used for Scheduling?

Architectural signatures of threads' are the basis on which threads' performance on each type of core is estimated at runtime. To accomplish this, in the paper the authors consider a hypothetical completion time of some constant number of instructions. Two parts of the instructions execution time has been considered separately- execution time (amount of time it takes to execute the instruction) and stall time (number of cycles due to last level cache misses and accessing main memory).

To determine stall time, latency of main memory access time and cache miss rate are required. Memory access latency can be discoverable by the OS and cache miss rate can be obtained from signature. The sum of the execution time and stall time gives an abstract completion time metric. Ratio of completion time calculated from different types of cores is used for actual scheduling.

### D. The Scheduler

Two key abstractions have been introduced in the paper-processor class and CPU partitions. A processor class has a distinct set of characteristics, has distinct set of attributes. For example, classes may be varied by clock frequency, cache hierarchy; execution cores and a system to be heterogeneous there should be at least two different types of core. A processor class can have large number of cores, resulting difficult operation of load balancing and accounting. So, another key abstraction CPU partition technique may be used here. Processors are grouped together in partitions, each processor will be in exactly one processor class where each processor class may consist of one or more partitions. During normal operation rather then the processor class, partition can be wildest locking scope. The number of threads that are currently running or ready to run that means runnable threads are kept tracking by counter in each partition. Counter of runnable threads in each partition is updated in real time and it has to be fully synchronized, as this is partition wide contention point.

When threads come to the system, using signature they estimate their performance on each processor class according to the characteristics of the class. There will be estimation per processor class. These values can be calculated once and used throughout the whole life cycle of the threads. These base ratings indicate the expectancy of performance if threads map themselves to a core in processor class for exclusive use.

The threads assign themselves to a partition using a process called regular assignment. Thread goes through all partitions and estimates its performance on that partition using base ratings and the number of runnable threads per core of that partition. Thread assigns itself to a partition where it gets higher estimated performance. With respect to the number of partitions, this type of assignment has linear complexity. When large processor class requires to be partitioned then a balance should be required between the number of partitions and number of cores in each partition. Assignment is a repeated process. It is done in a *refresh* manner. Every time a thread accumulates certain time of CPU time on its current partition or when the number of partition changes the current partition becomes non-optimal and assignment is done again.

Load balancing is an important issue while scheduling. Scheduler should not work in a manner such that some cores becomes heavily loaded with threads while other some cores run under utilization. To emphasize scalability this paper uses regular operating system load balancing and core assignment within partitions policies. Moreover, there is no direct load balancing between two partitions. Threads themselves goes for balanced load distribution, resulting more powerful partitions receiving higher loads. A situation may occur when a thread is waiting in a queue when there exist an idle core in the system. So, if there is any partition that is not fully loaded, threads are restricted to go for fully loaded or overloaded partition.

The greedy approach has a possibility to be generating local sub-optimal assignment. In this case increased optimization can be obtained only through co-operative action between two threads. A mechanism defined in the paper as *optimistic assignment* resolves this swapping issue between threads. During a refresh a thread can select this rather than choosing regular assignment, if it fails to choose a good target partition. It is the responsibility of initiator to find a partner in target partition and swap it. The initiator only can trigger this switching if it becomes obvious that the swapping will increase the performance of the thread and as well as the system performance. Using the base performance rating of initiator and the potential partner this can be done. If there is lot of partitions or if target partition has lot of threads then searching for a partner may be slow. So, exhaustive search will not be a good approach and randomized search with limited probing will be a good choice here. Optimistic rebinding is especially significant when partition underload protection mechanism starts. Situation like this regular assignment to any partitions except that are underloaded is not permitted. But rebind optimistically to any partition is permitted, even for those that are not underloaded. The reason of this, the fact that, swapping threads can't create more load imbalance than which already exists.

With the use of partition scheme, scheduler avoids global locks during scheduling. When doing a refresh threads can lock partition at a time to read the runnable threads counter or to migrate threads between partitions. If read/ write locks are used, the pressure on contention point can be reduced.

*E. Evaluation*

Two machines are used for experiments, Intel Xeon X5365 server with four dual core packages and an AMD Opteron 8356 with four quad core chips. Setting cores to run at different speeds using DVFS creates heterogeneity in the experiment. SPEC CPU200 suite is used for evaluation. For most of tests two categories of workloads are used, highly heterogeneous (HH) and moderately heterogeneous (MH). Highly heterogeneous workload consists of pair of CPU-bound benchmarks and a pair of memory-bound benchmarks. On the other hand, second category benchmarks represent whole spectrum of memory-boundedness.

HASS completion times have been compared with two composite metrics, default metric and ideal round robin metric. Default metric is weighted average of completion time of benchmarks when all of them are bound to specific type of core. This is simply expected execution time if a benchmark is randomly binds to a core and never switches and it gives a good assumption of how default scheduler works. On the other hand the second metric, ideal round robin metric, is calculated by combining completion time on all types of core types. Ideal round robin metric resembles the hypothetical scheduler that is perfectly fair compare to default scheduler.

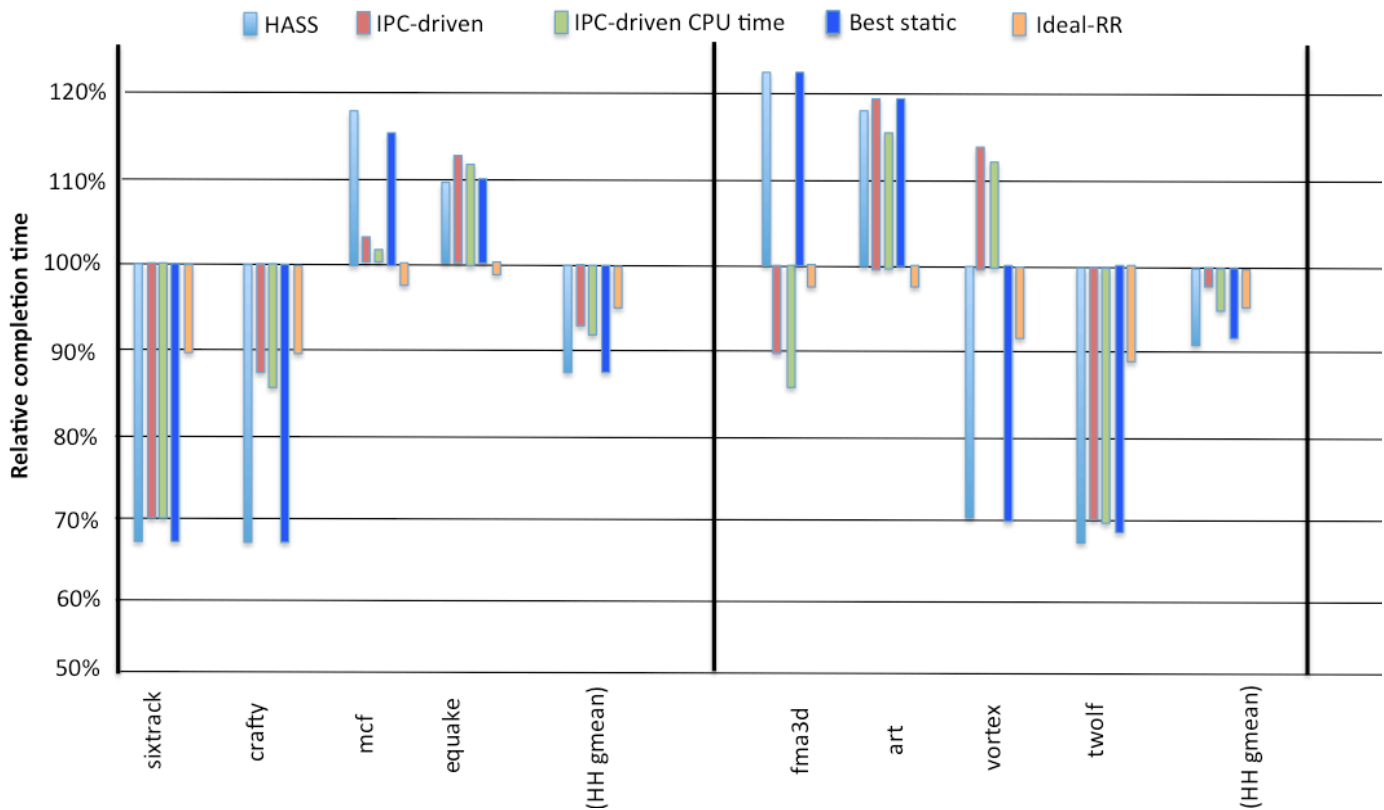

From the experiment it shows that (from figure 1, redrawn from figure 2 of [29]) HASS scheduler performs well in case of highly heterogeneous workload with an average speedup 13% on AMD machine, which is within 0.5% of the speedup on the best static assignment. Although in terms of moderately heterogeneous workload according to authors it is difficult to optimize.

The results of the experiment show that HASS is able to identify the difference among benchmarks in terms of architectural properties, able to map them in appropriate cores while scheduling, especially while considering highly heterogeneous workload. HASS also performs better than default metric even in homogeneous workload where performance improvement is difficult to obtain.

Scalability analysis of HASS scheduler shows how overhead scales up as complexity grows. The authors focus on part more specific to the algorithm, the time spend in partitioning. If set up complexity increases partition logic are supposed to take more time. But the result of the experiment shows that at 16 cores the overhead is insignificant, giving maximum overhead 0.06% of CPU time of a thread. So, partition assignment may not be a bottleneck, at least for medium size set up.

Summarizing all, HASS has several advantages. The first easily noticeable characteristics may be it is simple, so implementation is easy. HASS provides better scalability (al least medium size set up shows that), although there is no indication of how it will work for increased level of scalability. HASS provides support for short-lived threads, otherwise these threads would spend whole or majority of life cycle in performance monitoring phase, which is not optimal.

HASS has some limitations also. The paper discusses single threaded applications where there is only one signature per thread. Although, it has been mentioned that this can be extended to multithreaded application, it is difficult to accommodate varying input set. Varying input set can change application behavior and optimal thread to core mappings can be changed.

While developing application signature HASS requires cooperation from application development side, giving another limitation of the scheduler.

This paper indicates cycles penalty due to last level cache misses as constant, to say it simply assumes constant memory latency while using signature for scheduling. But the presence of non –uniform memory access can give wrong estimate in this case. Moreover, performance of different threads on different cores has been predicted using caching behavior and cores cache size and frequency, but accuracy of relative performance estimation for cores that differ in cache size is not done.

The biggest issue of HASS is that it is unaware of shared cache. HASS also behave unfair as it assigns jobs to fast core, which experiences most system speed up on those cores. But if a memory bounded job has higher priority the situation should not be like this.

Finally HASS does not aware of phase changes. An architectural signature persists for the lifetime of application, making HASS phase unaware.

V. TWO MORE EFFICIENCY SPECIALIZATION APPROACHES

A. Preliminary

The paper of Kumar et al [12] is one of the earliest and well-cited papers of single-ISA heterogeneous architectures. In this paper they have explored a new approach of multi-core architecture named CMP, which comprised of heterogeneous set of processor cores all of which can execute the same ISA. This paper has two major findings, which have influenced the later research of this area. They demonstrate that CMP approach provide significant performance advantages for a multiprogrammed workload compare to homogeneous chip multiprocessor. More specifically, they show that, a heterogeneous multiprocessor comprised of two cores gains as much as 63% performance improvement compare to equivalent area homogeneous multiprocessor. The second finding is, dynamic thread to core assignment policy plays a significant role regarding the performance gain. They have provided dynamic thread to core assignment policies that outperform the random scheduler and even beat the best static assignment. The policies dynamically determine the best core for threads either periodically or in response to triggering events. The best of these policies outperforms naive assignment by 31%.

B. Evaluation

The paper presents result using simulation approach. The model of the simulation consists of a number of chip multiprocessing configurations that is derived from combinations of processors from Alpha architecture family – EV5 (21164) and EV6 (21264). To support heterogeneous multicore architecture with multithreaded core they have considered EV6+, a hypothetical multithreaded version of EV6 processor. All cores are clocked 2.1 GHz frequencies. The workloads are constructed using SPEC2000 benchmarks. Half of the benchmarks are integer benchmark and half of them are floating-point benchmark. Moreover, half of them have large memory footprint and rest of half have memory footprint. SMTSIM, a cycle accurate execution-driven simulator that simulates an out-of-order, simultaneous multithreading processor [35], is used to simulate the benchmarks. Weighed speedup is taken as an evaluation metric in this paper. Weighted speedup here is measured as a ratio of the individual IPCs of the threads constituting a workload and the IPC of the threads on baseline configuration when running alone.

There may be two dimension of diversity in an application mix that can be exploited by heterogeneous architecture: diversity between application and diversity overtime within a single application. The paper separates these exploitation policies. In first part of simulation result, it shows the performance of static assignment of applications to cores. Although, after enter and exit of the jobs, the best assignment of jobs to cores will change. Scheduling policy for homogenous CMP configuration is straightforward- any core can be assigned to any core as long as core is available. In heterogeneous case the scheduling policy seeks to match the optimal static configuration as closely as possible. The scheduling policy has no L2 cache interaction when determining static assignment of workload to cores. So, the configuration that gives maximum weighted speedup can be found by running each job alone on each of the unique cores and use that as a guide to thread to core static assignment policy. Fig 2 of [12] shows the weighed speedup versus number of threads for homogenous (20 EV5, 4 EV6) and heterogeneous (5 EV5 and 3 EV6) configurations. The result shows that with a simple approach of static scheduling heterogeneous architecture gives more weighted speedup than homogenous architecture for most level of threading. The heterogeneous configuration seeks to combine the efficiency of both the powerful processors (EV6) and less powerful processors (EV5). When the thread number is low, threads can run in EV6 processor to get increased per thread performance. When thread number is high, the application can run on the added EV5 to give higher overall throughput.

From the result we can see that from 1 to 3 threads the heterogeneous configuration achieve same weighted speedup to the homogenous EV6 CMP configuration. When the number of thread is four, EV6 configuration shows a slight more weighted speedup to heterogeneous case. But for more than four threads the heterogonous configurations shows better performance.

Heterogeneous configuration performs 37% better with an average 26% improvement considering 1-20 thread comparing to homogeneous processor with 4 EV6 cores. It performs up to 2.3 times better with an average 23% improvement comparing to 20 EV5 cores.

The analysis described so far is inter-thread diversity, meaning diversity among threads and assigning threads to core considering the diversity. But the resource demand changes across the phase of an application. So, the best match to a thread to a core also changes across the various phase of application. The paper discusses some implementable heuristics that dynamically adjust thread to core assignment to improve performance. Heuristics are sampling based. A trigger is generated after some time intervals during the execution of a workload. This trigger initiates a sampling phase. In this sampling phase, the scheduler reconsider thread to core assignment policy and changes the cores of an application if performance gain is possible. Using hardware performance counter the dynamic execution profiles of the applications that are running, are collected. A new thread to core assignment can be done using dynamic execution profiles of the threads. This assignment is employed during a much longer phase of execution, the steady phase, which continues until the next trigger.

*C. Core sampling strategies*

In sampling phase and steady phase a large number of application-to-core assignment permutations are possible. The number of permutations can be pruned significantly by assuming that an application will never run on a less powerful core if that leave a more powerful core idle. Samples of the assignments can be selected depending on the level of interactions at the L2 cache level. The paper explains three strategies for sampling the assignment space.

*Sample-one* sampling strategy samples as many assignments as needed to run each thread once on each core. The underlying assumption is that single sample is accurate, regardless what other jobs are doing. The assignment is done, which maximizes weighted speedup assuming that future performance will be the same as one sample for each thread.

*Sample-avg* sampling strategy is based on the assumption that multiple samples are needed to get the average behavior of a job on each core. Samples are taken as many times as there are threads running. Samples are different from one another and taken in such a manner that at least two runs of each thread on each core type is possible. The assignment is done, which maximizes weighted speedup based on average performance of each thread on each core.

*Sample-sched* sampling strategy is based on the assumption that we know little about a particular assignment unless we actually run it. The sampling strategy samples a number of possible assignments and chooses one of the assignments it sampled.

From the result presented in the paper (fig [4] of [12]), we see that sample-sched strategy performs the best and sample-avg has also very similar performance compare to sample-sched. Also sample-one sampling strategy is not much worse. Another significant result we can see from the figure is that the intelligent assignment policies make distinguishable performance difference, which outperforms the random core assignment policy by up to 22%.

*D. Trigger Mechanisms*

Sampling of assignments has to deal with two major conflicting issues- minimizing sampling overhead and reacting quickly to changes in workload behavior. The paper proposed two trigger mechanisms, one based on periodic timer and other based on events indicating significant changes in performance.

Periodic timer is based on varying time between sampling phases, meaning the length of steady phase. If the steady state phase is small in length then a greater amount of time is spent on sampling phases, meaning that sampling overhead increases. Comparing the average weighted speedup obtained with steady-phase lengths between 31.25 million and 500 million cycles for the sample average strategy the paper shows that sampling frequency has a second-order impact on performance while steady state length of 125 million cycles perform best overall. So, optimal sampling period or frequency can be obtained from average phase length of applications, and the ratio of the lengths of the steady phase and sampling phase. The advantage of time-triggered sampling is – this is easy to implement. But it does not consider inter-thread or intra-thread diversity fully. If the phase lengths of different applications in the workload mix are different then a fixed sampling frequency is inadequate.

The second class of trigger mechanisms is based on the run length behavior of the workload. After monitoring run-time behavior of the workload if sufficient changes of the workload is detected then trigger is initiated. In *individual-event* trigger mechanism, if steady state IPC of a thread is changed more than 50% then sampling phase in triggered. In *global-event* trigger mechanism the absolute values of the percent changes in IPC for each application is taken as summation and a sampling phase is triggered when this value exceeds 100%. *Bounded-global-event* trigger mechanism modifies *global-event* trigger if more than 300 million cycles has elapsed since the last sampling phase. It does the modifying by initiating sampling phase. It also avoids sampling if the global event trigger occurs within 50 million cycles since the last sampling phase.

A comparison of these three event based trigger along with time-based trigger presented in the paper using a steady state length of 125 million cycles and sample-avg core sampling strategy. The result shows that event-based trigger outperforms the best timer-based trigger and the static assignment approach. So, event based triggers achieve the two goals described earlier- minimized sampling overhead and reacting quickly to workload changes.

*E. Preliminary*
In [15] Becchi et al argues that the benefits of heterogeneous CMPs are strengthened using dynamic assignment policy that means a runtime mechanism that observes the behavior of the threads and migrates thread between cores. This paper is one of the earliest and well-cited papers in this area, just like the paper of Kumar et al.

With a simulation approach they have shown that a dynamic assignment can outperform static assignment by 20%- 40% on average case and by as much as 80% in extreme case. In this paper, two dynamic assignments policies such as round robin and IPC driven are defined and they have been compared with static assignment policy. Using the simulation the paper shows that a heterogeneous system with dynamic assignment policies can exploit thread parallelism more efficiently than a homogenous and a heterogeneous system using static assignment policy.

In simulation, homogenous and heterogeneous configurations of EV5 (Alpha 21164) and EV6 (Alpha 21264) have been used. Homogenous and heterogeneous configurations used are-

Homogenous configurations: 4 EV6 or 20 EV5
Heterogeneous configurations: 5 EV5 and 3 EV6, 10 EV5 and 2 EV6, 15 EV5 and 1 EV6.

The workload has been constructed based on programs from SPEC 2000 benchmark suite. Among them, five are integer and six floating point. Workloads are constructed with randomization to reduce the sensitivity of the results to the particular set of programs simulated.

The evaluation metric of this simulation approach is the speedup of the CMP configuration over baseline performance of single EV6 core. The speedup defined as the ratio between the global instruction count and the execution time.

*F. Assignment Policies:*
Static Assignment:
Best static assignment for both homogeneous and heterogeneous system is NP hard. So, it is common that solutions proposed regarding this relies on heuristics, which gives sub-optimal solutions. In the simulation model, described in this paper, two static scheduling algorithms have been implemented to compare with dynamic assignment policies- random and pseudo best static assignment.

Random static assignment does not have any prior knowledge about the workload behavior. So, it assigns threads to processors in a random way. However, the assignment tries to maximize EV6, which mean it assigns thread to EV6 first. If system has more threads ready to run than cores then as soon as a core is available the assignment policy assigns the core to that thread.

Pseudo best assignment is based on the assumption that the runtime characteristics of the thread that are ready to execute are known beforehand.

Round robin dynamic assignment:
Static assignment has some drawbacks. It does not capture the phase behavior of the program. Static assignment let powerful EV6 cores to remain in idle state. If a powerful core becomes idle, it remains in that state unless some unassigned thread exists. If global IPC is a taken as a performance metric then the slow threads on EV5 cores can also reduce the overall system performance. Round robin dynamic policy tries to compensate the effects pose by the assignment policy. Threads are periodically assigned to processors in a round robin fashion ensures that available EV6 cores are equally shared among running programs. When one of EV6 core becomes idle and all threads are already assigned then threads from EV5 is migrated to EV6 for better utilization of EV6 and overall system.

IPC driven dynamic assignment:
In any instant of execution in homogeneous system, if threads can be executed in cores such that those cores maximizes overall system performance at that moment, then such assignment will be optimized the performance of homogeneous system. On other hand, the threads that get medium performance gain while running in fast core, may be ran on slow cores and migrated later when fast cores become idle. IPC driven dynamic assignment is based on this concept. As we are considering EV6 as fast processor and EV5 as slow processor, the ratio of the IPC on EV6 and EV5 processor may be use as the guide for assignment. Threads with higher IPC ratio may run in EV6 and threads with lower IPC ration may run in EV5. One important thing of this assignment policy is that assignment is based on IPC ratio. So, IPC values on both processors must be available in order to make assignment decision. Some learning mechanisms can be established to get such information.

*G. Evaluation Results*

Some of the key simulation results from the paper are presented below-

Round robin dynamic assignment policy performs better than pseudo optimal assignment policy on a 3EV6-5EV5 configuration. Even for the small number of threads all dynamic policy perform better than best static configuration on a 3EV6-5EV5 configuration. These results show us that for low number of threads dynamic assignment is better than ideal static assignment. For high degree of thread level parallelism, the presence of single EV6 guarantees better performance than the homogeneous case for fewer than 14 threads and a 2EV6-10EV5 configuration allows comparable speedup up to 30 threads. In comparing IPC driven dynamic policy and round robin assignment policy, simulation results shows that IPC driven performs better than round robin. The comparative performance gain increases with the increased number of threads exceeding the total number of cores.

VI. TLP SPECIALIZATION SCHEDULING ALGORITHM

In [28] Saez et al proposed a thread level parallelism specialization scheduling algorithm. Asymmetric multi processor system usually contains several fast and powerful cores and a large number of slower low-power cores. Fast cores are characterized by high clock frequency, complex out-of-order pipeline, and high power consumption. On the other hand, slow cores are characterized by low clock frequency, simple pipeline and low power consumption. A large number of slow cores are good for running parallel applications. Small number of fast and complex cores are good for running applications that are single threaded and sequential, because their performance can not be improved by letting run across multiple simple cores. According to the paper, due to the performance-power trade off, it becomes that it is more efficient to run parallel application on a large number of simple cores than a small number of complex cores.

*A. TLP specialization on AMPs*

Catering to diversity of thread-level parallelism is one way to improve efficiency on AMPs. Application can be classified into two classes of categories in respect of diversity of thread-level parallelism. They are scalable parallel applications and sequential applications. Scalable parallel thread contains multiple threads in execution and increased number of threads means that execution time is reduced. On the other hand, sequential applications contain small number of threads and it is difficult to structure them in multithreaded environment such that efficiency is improved. Other than parallel and sequential application there exist one kind of hybrid application where application may have phases of highly parallel execution intermixed with sequential phases.

Given two types of different type of workload, its obvious that we need different types of processing cores to achieve best trade-off in performance and energy consumption. For example, suppose we have four complex-powerful cores processor and sixteen simple but low power cores. Lets suppose that complex cores give twice performance improvement than slow cores. We have scalable parallel applications that we can choose either to run in complex powerful cores or simple- slow power cores. If we run the scalable parallel application in complex cores then threads in complex core can run twice faster than thread running in slow cores, but we have only four complex cores compare to sixteen simple cores. But as parallel application is scalable, using additional threads and running them in simple cores gives as much as twice system performance improvement compare to running them in complex cores. On the other hand, consider we have sequential application that we choose to run on either complex cores or simple cores. Sequential application cannot increase its performance by using additional threads. It turns out that if we run single threaded sequential application in slow cores then it will run twice slower than if we run this in complex cores.

These example show that depending upon the parallelism of the application, we require different types of cores to obtain optimal performance-per-watt ratio. Asymmetric multicore system resolves this issue by providing different types of cores.

*B. TLP based scheduling algorithm*

It is the task of the software to employ TLP specialization scheduling policies to get the benefit on asymmetric cores; specialization on AMP systems will not be delivered by the hardware. A thread scheduler must aware of asymmetry of the cores and has to map the application to cores according to this.

The idea or scheduling algorithm called as parallelism aware (PA) scheduling algorithm described in the paper is very simple. The scheduler assigns threads running highly parallel code on slow cores. It assigns threads that are sequential or sequential phases of parallel application to fast cores.

Figure [2] (redrawn from figure 2 of [28]) shows an illustration of how a PA scheduler would accelerate a parallel application limited by a sequential bottleneck on AMP processor.

The authors present a simulation done in OpenSolaris system comparing the performance of a number of parallel applications on an AMP system using their proposed PA scheduler and default symmetry unaware scheduler. To emulate the AMP they have used AMD Opteron with 16 cores. They fast cores have clock frequency of 2.3GHz and slow cores have low clock frequency- 1.15GHz. They have used 4 cores as fast cores and 12 cores as slow cores. Several benchmark suites such as SPEC OpenMP 2001, PARSEC, MineBench, and NAS have been used in simulation.

The simulation result is presented in figure [3] taken from figure [3] of [28]. The result shows that application with around 40%-60% sequential phases gives a performance improvement of up to 26% compare to the scheduler that is asymmetry unaware. Although applications with small sequential phases do not give any speed up using PA scheduling algorithm.

*C. Challenges involving PA scheduling:*
There are two significant challenges that require overcoming while implementing PA scheduling. The first challenge is effectively detecting parallel and sequential phases of an application. The second one is, the migration overhead due to threads cross core migration. The authors explain heuristics to overcome these two challenges. To detect sequential and parallel phase of application runnable thread count can be used as a heuristic. The application that uses thus this application is in parallel phase. On the other hand, the application with only one thread that means runnable thread count one, is in sequential phase. The good thing is that, in multithreading environment operating system has the knowledge of runnable thread count. So, by using runnable thread count operating system can identify parallel and sequential phases of application.

If an application runs on non-scalable code while still significant amount of runnable thread, then runnable thread count for detection of sequential phases may not work. Consider one scenario where an application may be affected by an external bottleneck due to result of memory bandwidth contention. As the memory bus is saturated the additional threads will not improve efficiency here. A solution may here is reduction of threads used in an application where it application runs in peak efficiency. Feedback-driven threading, a technique described by Suleman et al [36] can be used to dynamically determine the optimal thread count for parallel application.

In another scenario, internal scalability may be a bottleneck for an application. Load imbalance and neck may happen where one thread does more work than other and where one thread may execute the code in a critical section while other threads wait. If a thread is in block state then runnable thread count is reduced and operating system acknowledged

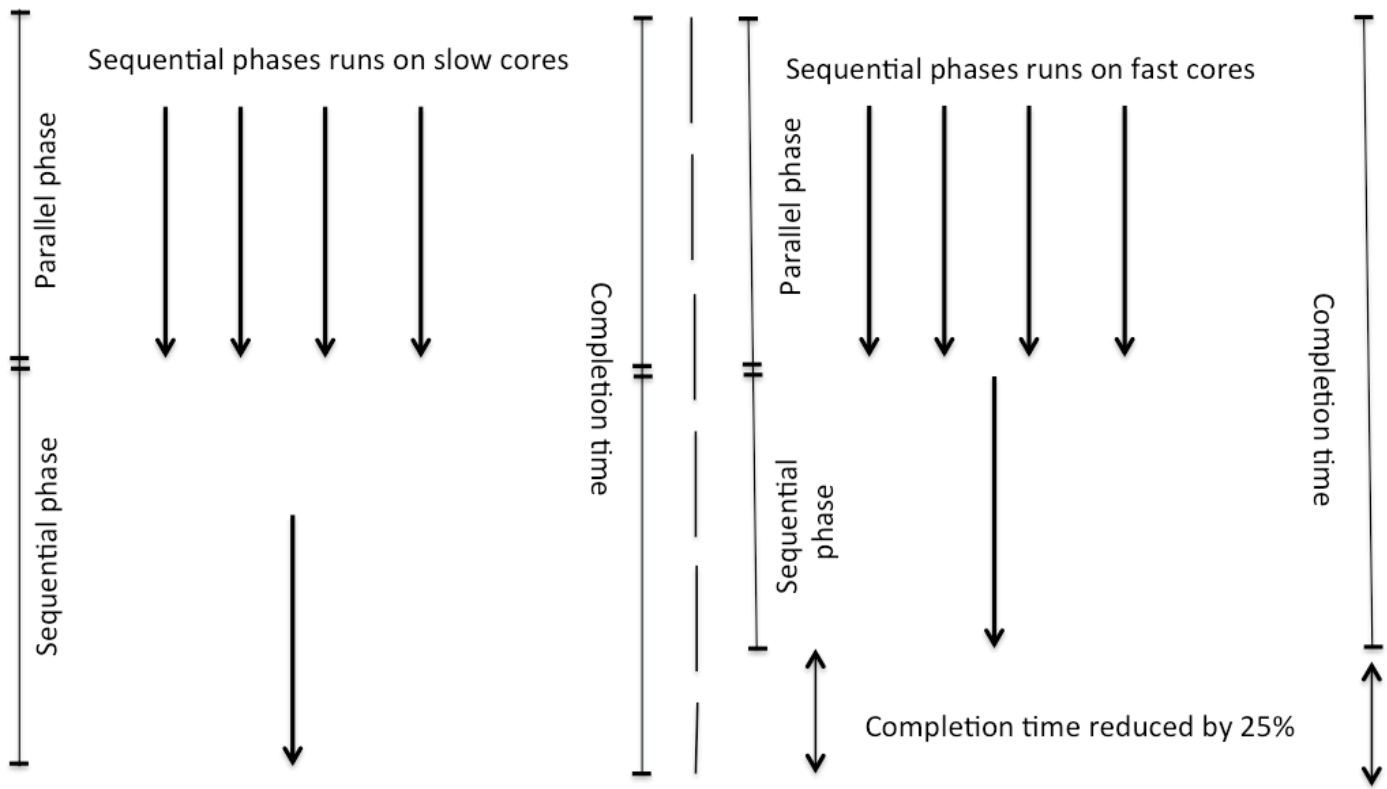

Figure 2 [fig 2 of [28]]: An illustration of how PA scheduler would accelerate a parallel application limited by a sequential bottleneck on an AMP processor

large number of threads should have high runnable counts, this reduction of runnable thread. But, if thread is in busy-wait state, sequential phases may not be exposed to operating system. The paper shows that for application with large sequential phases, performance improvement may be up to 40% if sequential phases are exposed to the scheduler via adaptive synchronization.

Another challenge of implementing PA scheduler is avoiding the overhead of thread migration in cores. Symmetry aware scheduling algorithms rely on cross core migration policy to get benefits from AMP structure. So, PA algorithm must migrate a thread from simple core to the same memory domain like several slow cores then situation becomes completely in favor. So, scheduler does not require migrating thread between cross memory domain. Scheduler will migrate thread from slow to fast core in the same memory domain, giving the opportunity to reuse in LLC. PA scheduler is topology aware. So, it tries to avoid cross-memory-domain thread migration whenever possible. Migration relation performance overhead of PA scheduler can be obtained from comparing the performance of applications under PA scheduler and default scheduler. Migration overhead in this case is equivalent to performance complex core if it finds that thread is executing in sequential phase. Although migration is an essential feature of asymmetry aware algorithms, migration may become an overhead if it appears that migration is expensive.

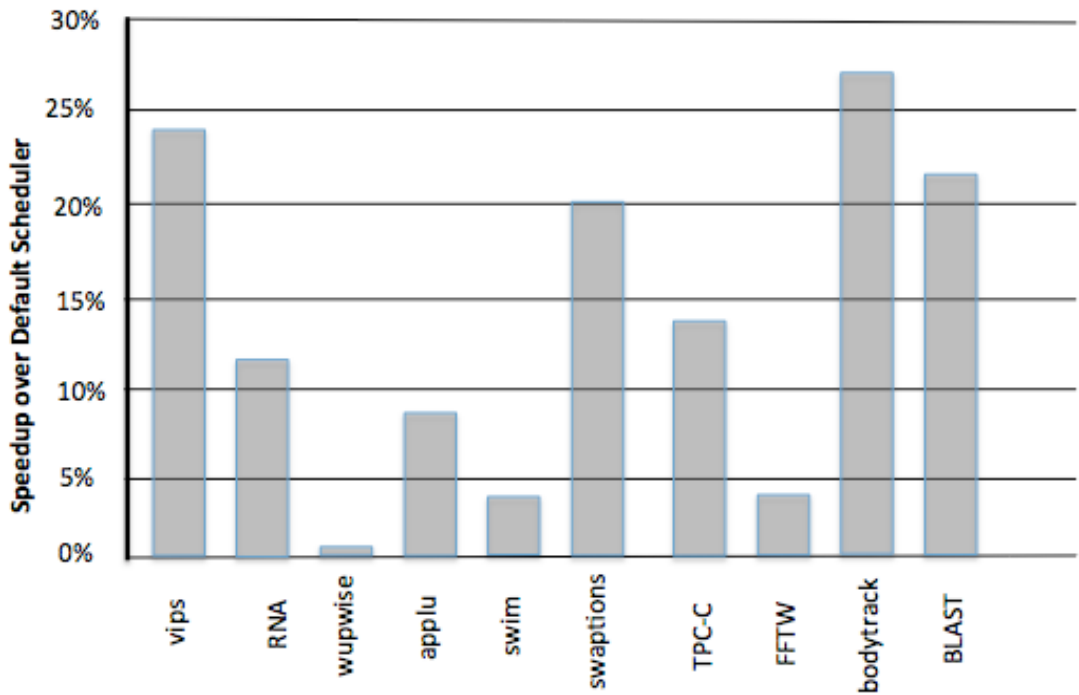

Fig 3 [figure [3] of [28]]: Speedup achieved with PA algorithm over the asymmetry-agnostic default scheduler on an emulated AMP system

Typically an AMP system consists of several memory domains. A memory domain can be defined as the collection of cores that share the same last level cache (LLC). LLC is the last hope for the processor to get the data without wasting too much CPU cycles. If a requested data is not in LLC then processor has to fetch the data from main memory, which takes hundreds of CPU cycles and slows down the operation of CPU. But if the data is in LLC then processor does not have to waste the CPU cycles. While thread migration LLC misses becomes a major issue. If fast cores in AMP system are located in different memory domains than slow cores, then scheduler's migration of a thread from slow core to fast core may impact severely. The thread loses the data that was in the LLC of slow memory domain, so data has to be fetched from main memory. But if in an AMP system has architecture such that fast cores are in degradation. Comparing performance overhead relative to default scheduler for migration-unfriendly and migration friendly topology this paper shows that performance overhead becomes significant in migration-unfriendly system, but if topology aware scheduler is used in migration friendly system then the overhead becomes negligible.

To summarize all these things the paper concludes that for parallel applications limited by sequential phases parallelism- aware algorithm can produce significant performance improvement on asymmetric hardware. Configuring synchronization to detect sequential phases of thread in a major way to success. AMP system should be designed in a way such that memory domains of some slow cores and fast cores become same, meaning fast cores and some slow cores should be in same memory domain to avoid cross memory domain thread migration. Topology aware scheduler should be used to migrate thread to reduce cross memory domain thread migration.

## VII. ALGORITHM EXPLOITING EFFICIENCY AND TLP

So far in this paper, we have seen two types of specialization scheduling algorithms: efficiency specialization, thread level parallelism (TLP) specialization. Efficiency specialization scheduling algorithms get improved efficiency by assigning CPU-intensive threads to fast cores. On the other hand, TLP specialization algorithms get improved efficiency by assigning sequential applications and sequential phases of parallel applications to execute on fast cores. These two types of schedulers work effectively for different types of workload. For single threaded application efficiency specialization algorithms deliver greater benefit and for parallel applications TLP specialization algorithm proved to be effective. Now question arises, is there any scheduling algorithm that targets both efficiency specialization and TLP specialization? How good the algorithm will be?

In [16] Saez et al proposed a scheduler named CAMP, a comprehensive AMP scheduling algorithm, which gives both types of specialization. To determine which threads are best candidates to run on fast cores they introduce a new metric, utility factor (UF). For each application, this utility factor considering application efficiency and TLP gives a single value that indicates how much an application performance will be improved if all of its threads are allowed to run on fast cores. This paper also introduces a new way to determine the efficiency of a specific thread running on a fast core. Typically, speed up factor gives the measurement of efficiency where speed up factor is the relative improved running time of a thread from fast core to slow core. This paper introduces a new method to estimate speedup factor by measuring last level cache misses. This speedup factor categorizes application low, medium and high classes according to their efficiency.

### A. Utility Factor

Let a system with $N_{FC}$ fast cores. Utility factor (UF) metric can be calculated from the application speed up if $N_{FC}$ of the threads of the application are placed to fast cores and remaining of the threads are placed in slow cores, comparing to placing all the threads in slow core.

Speedup= $T_{base}/ T_{alt}$
$T_{base}$ = Completion time of the application in base configuration where only slow cores are used.
$T_{alt}$ = Completion time of alternative configuration where both slow and fast cores are used.

UF formula can be expressed as following equation-

$$UF = SF_{app} / MAX(1, N_{THREADS} - (N_{FC}-1)^2)$$

$N_{THREADS}$ = Number of thread in the application.
$SF_{app}$ = Average speed up factor of the application's thread when running on fast core relative to a slow core.

The authors make two assumptions in constructing the utility factor-

1. The fast cores will be only allowed to use by the threads of target application for which UF is estimated.
2. The number of slow cores should be greater than the number of threads in application.

Using this model scheduler can estimate the utility factor for each application. The higher the utility factor of an application the more benefit can be got running the threads of application in fast cores. So, the scheduler will assign the application with highest utility factor to the fast cores.

### B. CAMP Scheduler- Algorithm

Utility factors of the threads give the opportunity to CAMP scheduler to assign them in different cores. Threads are categorized in three classes according to utility factor: LOW, MEDIUM and HIGH. So, the threads with relatively very close utility factor will be in the same class and class will also allow reducing any inaccuracies in estimation of SF used in calculating utility factor.

Threads those are in HIGH class will run on fast cores. If the number of threads in HIGH class is greater than the number of fast cores then the cores will be shared among these threads in a round robin fashion. If all the threads in HIGH class are running in fast cores and still there are some cores remain then threads from the MEDIUM class will be allowed to execute on fast cores. If there are no threads in MEDIUM class then threads from LOW class are allowed to execute on fast cores. But unlike the threads of HIGH class, the threads of MEDIUM and LOW class will not share the fast cores. Sharing of cores implies cross-cores migration and performance can be severely hurt by this. For example, memory extensive threads they cross core migration may cause last level cache misses and processor has to fetch the data from main memory in this case.

There will be a special class SEQUENTIAL_BOOSTED for the parallel applications executing a sequential phase. These threads will get the highest privilege to run on the fast cores. SEQUENTIAL_BOOSTED class will be comprised with threads having high utility factor value. But, in spite of having sequential phases the medium and low utility factor valued threads will be in regular class. To prevent the effect of threads to monopolizing the fast cores amp_boost_ticks, a configurable parameter is used. The thread will be in SEQUENTIAL_BOOSTED class during the duration amp_boost_ticks. After that threads will be downgraded to their regular class according to utility factor.

CAMP relies on two utility thresholds, lower and upper to determine boundaries between LOW, MEDIUM and HIGH utility classes. Lower threshold is used to denote the boundary between LOW and MEDIUM classes and upper threshold is used to denote the boundary between MEDIUM and HIGH classes. CAMP dynamically selects which utility threshold to use based on system workload. There are two pairs of utility thresholds used. One threshold is when only single threaded application runs in the system and the other is when at least one multi-threaded application is running on the system.

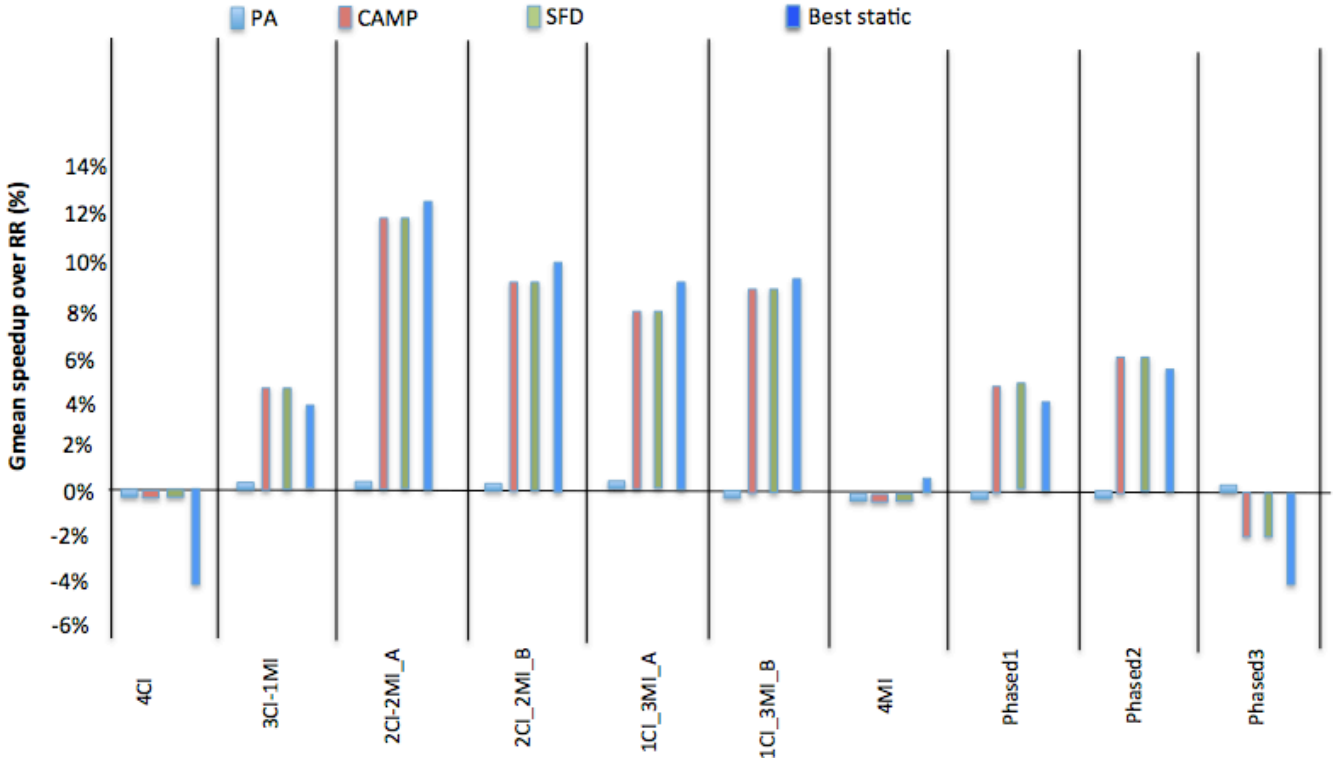

Figure 4 [fig 3 of [16]]: Speed up of PA, SFD, CAMP and Best Static schedulers when running single-threaded workloads on the 2FC- 2SC AMP platform

*C. Experimental results and analysis*

The authors present experiments and analysis to evaluate the CAMP scheduler. They have evaluated the accuracy of SF estimation, evaluated the method for single threaded application and presented aggregate results for all workloads comparing with other scheduler like parallelism aware (PA) scheduler and speedup factor driven (SFD) scheduler. They have used AMD Opteron system with four quad core CPUs. The system has a Non uniform memory architecture with sixteen cores. Each core can at a range of frequencies from 1.15 GHz to 2.3 GHz. As fast cores and slow cores are concerned, they have configured some cores to run at 2.3 GHz, which are fast cores and others to run at 1.15 GHz, which are slow cores. Three AMP configurations have been used in the experiments, one fast core and twelve slow cores (1FC-12SC), four fast cores and twelve slow cores (4FC-12SC), two fast cores and two slow cores (2FC-2SC).

They have constituted workloads taking applications from several benchmarks- SPEC OMP 2001, SPEC 2006, Minebench suites, BLAST and FFT-W.

To measure the accuracy of SF, they have compared the estimated SF to the actual SF for all applications in SPEC CPU2006. Actual speed up is the improvement of the running time of the application when it runs in fast cores relative to slow cores. Estimated SF in the paper is the average last level cache misses throughout the entire run of the application. Results of the paper show that the estimation of SF is accurate for CPU intensive application but less accurate for medium application.

Typically, efficiency specialization algorithm like SFD targets single threaded application to get improved efficiency and thread level parallelism (TLP) based algorithm like PA targets multi-threaded application. So, the paper works on both single thread and multi thread applications as it claims to get improved efficiency in both cases.

To evaluate CAMP scheduler speed up compare to other schedulers for single threaded applications, seven applications from SPEC CPU 2006 suite is chosen and ten workloads has been constructed. The workloads have variety; some of them are either memory-intensive or CPU intensive, others have different phases across application.

Figure 4 (redrawn Figure 3 of [16]) shows the speed up of PA, CAMP, SFD and Best Static scheduler when running single threaded workloads on the 2FC-2SC (two fast cores, two slow cores) AMD platform. Best static assignment ensures application with highest overall ratios run on fast cores. From the results it show that parallelism aware (PA) scheduler behaves like round robin scheduler. PA is unaware about the efficiency of threads. So, it assigns all the threads in HIGH utility class and assigns them in round robin fashion to fast cores. As UF=SF, CAMP and SFD perform similarly for this type of workload. From the figure we see that for most categories of workload CAMP and SFD efficiently distinguish CPU intensive and Memory intensive application and maps them in appropriate cores that is very closer to Best Static.

achieves higher speed up compare to PA and SFD scheduler across wide variety of workloads, which is a major contribution of the paper.

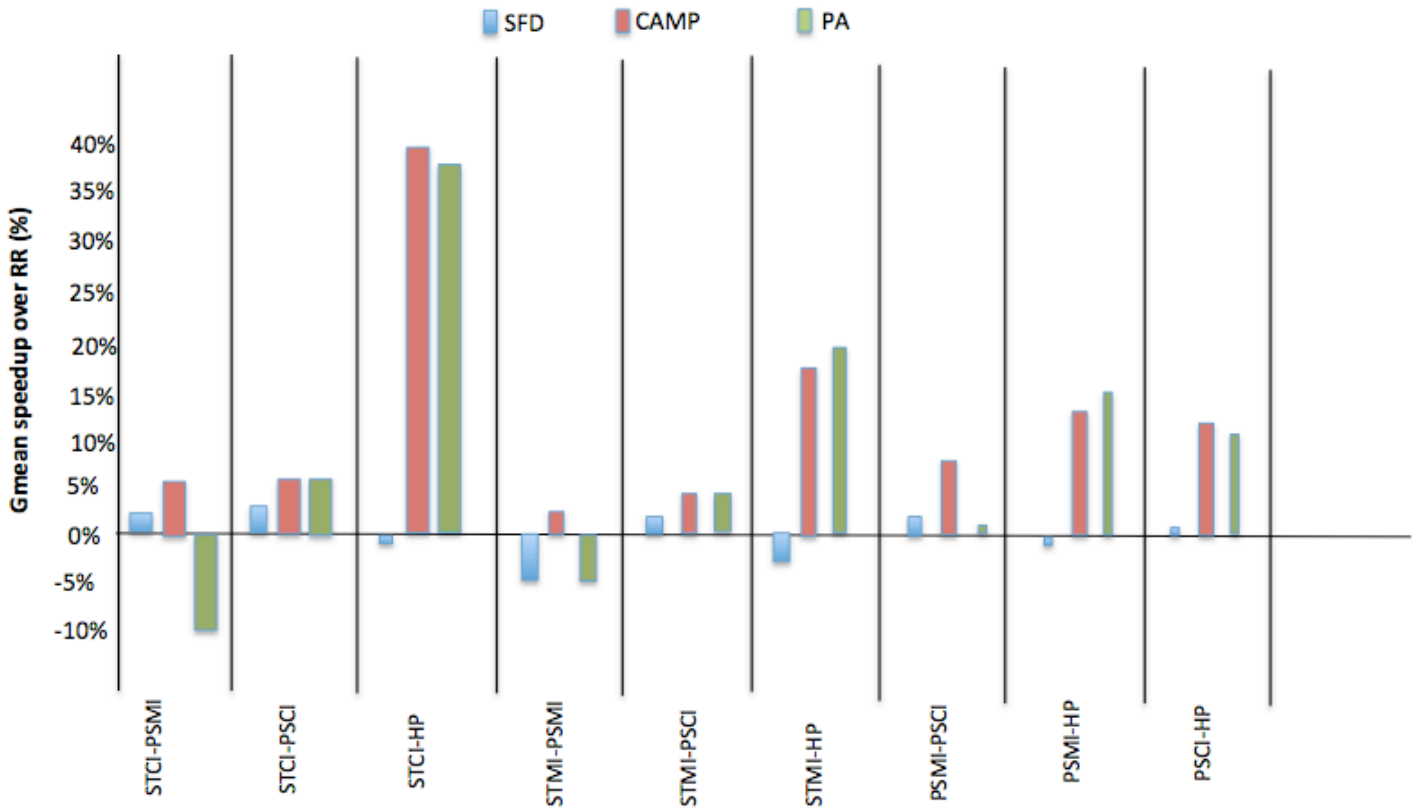

Figure 5 [fig 4 of [16]]: Gmean speed up of SFD, PA, CAMP schedulers when running multi-threaded workloads on AMD platform

For single threaded and multi-threaded applications, depending upon parallelism the paper categorizes application into three groups –
- Highly parallel application (HP)
- Partially sequential applications (PS)
- Single threaded application (ST)

Considering CPU intensity and memory intensity each group is again categorized into memory-intensive (MI) and CPU intensive (CI) classes. So, there is six application classes are possible, among which HPCI and HPMI class represents highly parallel applications, CPU intensive and memory intensive, respectively. PSCI and PSMI classes represent partially sequential applications. STCI and STMI classes represent single threaded application. The authors then constructs nine workload, where each workload is a pair two classes described earlier, like STCI-PSMI, STMI-PSCI.

Figure 5 (Figure 4 of [16]) shows geometric mean speedup of the three schedulers SFD, PA and CAMP normalized to RR for the workload described earlier. The nine workloads are certainly diversified, they different in terms of parallel phase, sequential phase, single thread, memory or CPU intensity. But the result shows that CAMP scheduler

## VIII. CONCLUSION

In this paper, taxonomies of scheduling algorithms for asymmetric multicore architecture have been discussed and some representative algorithms from each class have been discussed. While scheduling, schedulers target some sort of specialization. Some algorithms target efficiency, some algorithms target thread level parallelism and others target both. Efficiency specialization algorithms try to get better utilization and performance by assigning CPU intensive threads to powerful cores and TLP specialization algorithms assigns sequential applications and sequential phase of parallel application to powerful cores. The efficiency of scheduling algorithms depends upon the type of the workload. For single threaded workload efficiency specialization algorithm shows better performance, but if parallel applications is present then TLP specialization algorithms show better performance.

16